\pdfoutput=1
\documentclass[12pt,a4paper]{article}
\usepackage[T1]{fontenc}
\usepackage{ifthen} % for conditional statements
\usepackage{comment}
\usepackage{multirow}
\newboolean{pdflatex}
\setboolean{pdflatex}{true} % False for eps figures 

\newboolean{articletitles}
\setboolean{articletitles}{true} % False removes titles in references

\newboolean{uprightparticles}
\setboolean{uprightparticles}{false} %True for upright particle symbols

\def\paperauthors{LHCb collaboration} % Leave as is for PAPER, CONF and FIGURE
\def\paperasciititle{Observation of Bc+ -> D h+ h- decays} % Set ASCII title here !! MAKE sure it's only ASCII characters !! 
\def\papertitle{Observation of  $B_c^+ \to D h^+ h^-$ decays} % D_{(s)}^{(*)+}
\def\paperkeywords{{High Energy Physics}, {LHCb}} % Comma separated list
\def\papercopyright{\the\year\ CERN for the benefit of the LHCb collaboration} % new since 9/Apr/2018
\def\paperlicence{CC BY 4.0 licence}
\def\paperlicenceurl{https://creativecommons.org/licenses/by/4.0/}

\usepackage[top=1in, bottom=1.25in, left=1in, right=1in]{geometry}

\usepackage{microtype}
\usepackage{lineno}  % for line numbering during review
\usepackage{xspace} % To avoid problems with missing or double spaces after
\usepackage{caption} %these three command get the figure and table captions automatically small

\usepackage{graphicx}  % to include figures (can also use other packages)
\usepackage{color}
\usepackage{colortbl}
\graphicspath{{./figs/}} % Make Latex search fig subdir for figures
\usepackage{amsmath} % Adds a large collection of math symbols
\usepackage{amssymb}
\usepackage{amsfonts}
\usepackage{upgreek} % Adds in support for greek letters in roman typeset

\newcommand*\patchAmsMathEnvironmentForLineno[1]{%
\expandafter\let\csname old#1\expandafter\endcsname\csname #1\endcsname
\expandafter\let\csname oldend#1\expandafter\endcsname\csname
end#1\endcsname
 \renewenvironment{#1}%
   {\linenomath\csname old#1\endcsname}%
   {\csname oldend#1\endcsname\endlinenomath}%
}
\newcommand*\patchBothAmsMathEnvironmentsForLineno[1]{%
  \patchAmsMathEnvironmentForLineno{#1}%
  \patchAmsMathEnvironmentForLineno{#1*}%
}
\AtBeginDocument{%
\patchBothAmsMathEnvironmentsForLineno{equation}%
\patchBothAmsMathEnvironmentsForLineno{align}%
\patchBothAmsMathEnvironmentsForLineno{flalign}%
\patchBothAmsMathEnvironmentsForLineno{alignat}%
\patchBothAmsMathEnvironmentsForLineno{gather}%
\patchBothAmsMathEnvironmentsForLineno{multline}%
\patchBothAmsMathEnvironmentsForLineno{eqnarray}%
}

\usepackage{hyperxmp}

\usepackage[pdftex,
            pdfauthor={\paperauthors},
            pdftitle={\paperasciititle},
            pdfkeywords={\paperkeywords},
            pdfcopyright={Copyright (C) \papercopyright},
            pdflicenseurl={\paperlicenceurl}]{hyperref}
\usepackage[colorinlistoftodos,textsize=scriptsize]{todonotes}

\usepackage[bottom,flushmargin,hang,multiple]{footmisc}

\usepackage[all]{hypcap} % Internal hyperlinks to floats.

\usepackage{xspace} 
\usepackage{upgreek}

\def\lhcb   {\mbox{LHCb}\xspace}

\def\MagUp {\mbox{\em Mag\kern -0.05em Up}\xspace}

\ifthenelse{\boolean{uprightparticles}}%
{

 \def\Ppi         {\ensuremath{\uppi}\xspace}

 \def\PDelta      {\ensuremath{\Delta}\xspace}                 
 \def\PXi         {\ensuremath{\Xi}\xspace}                 
 \def\PLambda     {\ensuremath{\Lambda}\xspace}                 
 \def\PSigma      {\ensuremath{\Sigma}\xspace}                 
 \def\POmega      {\ensuremath{\Omega}\xspace}                 
 \def\PUpsilon    {\ensuremath{\Upsilon}\xspace}
 \let\oldPi\Pi
 \def\PPi         {\ensuremath{\oldPi}\xspace}

 \def\PB      {\ensuremath{\mathrm{B}}\xspace}                 
                  
 \def\PD      {\ensuremath{\mathrm{D}}\xspace}

 \def\PK      {\ensuremath{\mathrm{K}}\xspace}

 \def\Pb      {\ensuremath{\mathrm{b}}\xspace}                 
 \def\Pc      {\ensuremath{\mathrm{c}}\xspace}                 
 \def\Pd      {\ensuremath{\mathrm{d}}\xspace}                 
 \def\Pe      {\ensuremath{\mathrm{e}}\xspace}

 \def\Pi      {\ensuremath{\mathrm{i}}\xspace}

 \def\Ps      {\ensuremath{\mathrm{s}}\xspace}                 
                  
 \def\Pu      {\ensuremath{\mathrm{u}}\xspace}

 \def\thebaroffset{0.0em}
}
{

 \def\Ppi         {\ensuremath{\pi}\xspace}

 \mathchardef\PDelta="7101
 \mathchardef\PXi="7104
 \mathchardef\PLambda="7103
 \mathchardef\PSigma="7106
 \mathchardef\POmega="710A
 \mathchardef\PUpsilon="7107
 \mathchardef\PPi="7105
                  
 \def\PB      {\ensuremath{B}\xspace}                 
                  
 \def\PD      {\ensuremath{D}\xspace}

 \def\PK      {\ensuremath{K}\xspace}

 \def\Pb      {\ensuremath{b}\xspace}                 
 \def\Pc      {\ensuremath{c}\xspace}                 
 \def\Pd      {\ensuremath{d}\xspace}                 
 \def\Pe      {\ensuremath{e}\xspace}

 \def\Pi      {\ensuremath{i}\xspace}

 \def\Ps      {\ensuremath{s}\xspace}                 
                  
 \def\Pu      {\ensuremath{u}\xspace}

 \def\thebaroffset{0.18em}
}
\newcommand{\offsetoverline}[2][\thebaroffset]{\kern #1\overline{\kern -#1 #2}}%

\makeatletter
\ifcase \@ptsize \relax% 10pt
  \newcommand{\miniscule}{\@setfontsize\miniscule{4}{5}}% \tiny: 5/6
\or% 11pt
  \newcommand{\miniscule}{\@setfontsize\miniscule{5}{6}}% \tiny: 6/7
\or% 12pt
  \newcommand{\miniscule}{\@setfontsize\miniscule{5}{6}}% \tiny: 6/7
\fi
\makeatother

\DeclareRobustCommand{\optbar}[1]{\shortstack{{\miniscule (\rule[.5ex]{1.25em}{.18mm})}
  \\ [-.7ex] $#1$}}

\def\epem       {{\ensuremath{\Pe^+\Pe^-}}\xspace}
\def\uquark    {{\ensuremath{\Pu}}\xspace}
\def\uquarkbar {{\ensuremath{\overline \uquark}}\xspace}

\def\dquark    {{\ensuremath{\Pd}}\xspace}
\def\dquarkbar {{\ensuremath{\overline \dquark}}\xspace}

\def\squark    {{\ensuremath{\Ps}}\xspace}
\def\squarkbar {{\ensuremath{\overline \squark}}\xspace}

\def\cquark    {{\ensuremath{\Pc}}\xspace}

\def\bquark    {{\ensuremath{\Pb}}\xspace}
\def\bquarkbar {{\ensuremath{\overline \bquark}}\xspace}

\def\pion   {{\ensuremath{\Ppi}}\xspace}

\def\pip    {{\ensuremath{\pion^+}}\xspace}
\def\pim    {{\ensuremath{\pion^-}}\xspace}

\def\kaon    {{\ensuremath{\PK}}\xspace}
\def\KorKbar {\kern \thebaroffset\optbar{\kern -\thebaroffset \PK}{}\xspace}

\def\Kp      {{\ensuremath{\kaon^+}}\xspace}
\def\Km      {{\ensuremath{\kaon^-}}\xspace}

\def\Kstarz  {{\ensuremath{\kaon^{*0}}}\xspace}

\def\Kstar   {{\ensuremath{\kaon^*}}\xspace}

\def\D       {{\ensuremath{\PD}}\xspace}

\def\DorDbar {\kern \thebaroffset\optbar{\kern -\thebaroffset \PD}\xspace}
\def\Dz      {{\ensuremath{\D^0}}\xspace}

\def\Dp      {{\ensuremath{\D^+}}\xspace}
\def\Dm      {{\ensuremath{\D^-}}\xspace}

\def\DpDm    {\ensuremath{\Dp {\kern -0.16em \Dm}}\xspace}

\def\Dstarp  {{\ensuremath{\D^{*+}}}\xspace}

\def\Ds      {{\ensuremath{\D^+_\squark}}\xspace}
\def\Dsp     {{\ensuremath{\D^+_\squark}}\xspace}
\def\Dsm     {{\ensuremath{\D^-_\squark}}\xspace}

\def\Dss     {{\ensuremath{\D^{*+}_\squark}}\xspace}

\def\B       {{\ensuremath{\PB}}\xspace}

\def\BorBbar {\kern \thebaroffset\optbar{\kern -\thebaroffset \PB}\xspace}
\def\Bz      {{\ensuremath{\B^0}}\xspace}

\def\Bd      {{\ensuremath{\B^0}}\xspace}

\def\BdorBdbar {\kern \thebaroffset\optbar{\kern -\thebaroffset \Bd}\xspace}
\def\Bu      {{\ensuremath{\B^+}}\xspace}

\def\Bp      {{\ensuremath{\Bu}}\xspace}

\def\Bs      {{\ensuremath{\B^0_\squark}}\xspace}

\def\BsorBsbar {\kern \thebaroffset\optbar{\kern -\thebaroffset \Bs}\xspace}
\def\Bc      {{\ensuremath{\B_\cquark^+}}\xspace}
\def\Bcp     {{\ensuremath{\B_\cquark^+}}\xspace}
\def\Bcm     {{\ensuremath{\B_\cquark^-}}\xspace}

\def\Y#1S{\ensuremath{\PUpsilon{(#1S)}}\xspace}

\def\LorLbar     {\kern \thebaroffset\optbar{\kern -\thebaroffset \PLambda}\xspace}

\newcommand{\decay}[2]{\ensuremath{#1\!\to #2}\xspace} 

\def\to                 {\ensuremath{\rightarrow}\xspace}

\def\CP                {{\ensuremath{C\!P}}\xspace}

\def\AT#1     {\ensuremath{A_{\mathrm{T}}^{#1}}\xspace}           % 2

\def\C#1      {\ensuremath{\mathcal{C}_{#1}}\xspace}                       % 9
\def\Cp#1     {\ensuremath{\mathcal{C}_{#1}^{'}}\xspace}                    % 7
\def\Ceff#1   {\ensuremath{\mathcal{C}_{#1}^{\mathrm{(eff)}}}\xspace}        % 9  
\def\Cpeff#1  {\ensuremath{\mathcal{C}_{#1}^{'\mathrm{(eff)}}}\xspace}       % 7
\def\Ope#1    {\ensuremath{\mathcal{O}_{#1}}\xspace}                       % 2
\def\Opep#1   {\ensuremath{\mathcal{O}_{#1}^{'}}\xspace}                    % 7

\newcommand{\aunit}[1]{\ensuremath{\text{\,#1}}}       
\newcommand{\tev}{\aunit{Te\kern -0.1em V}\xspace}
\newcommand{\gev}{\aunit{Ge\kern -0.1em V}\xspace}
\newcommand{\mev}{\aunit{Me\kern -0.1em V}\xspace}
\newcommand{\kev}{\aunit{ke\kern -0.1em V}\xspace}
\newcommand{\ev}{\aunit{e\kern -0.1em V}\xspace}
 
\newcommand{\mevc}{\ensuremath{\aunit{Me\kern -0.1em V\!/}c}\xspace}
\newcommand{\gevc}{\ensuremath{\aunit{Ge\kern -0.1em V\!/}c}\xspace}
\newcommand{\mevcc}{\ensuremath{\aunit{Me\kern -0.1em V\!/}c^2}\xspace}
\newcommand{\gevcc}{\ensuremath{\aunit{Ge\kern -0.1em V\!/}c^2}\xspace}
\def\fb   {\ensuremath{\aunit{fb}}\xspace}
\def\invfb   {\ensuremath{\fb^{-1}}\xspace}

\def\gsim{{~\raise.15em\hbox{$>$}\kern-.85em
          \lower.35em\hbox{$\sim$}~}\xspace}
\def\lsim{{~\raise.15em\hbox{$<$}\kern-.85em
          \lower.35em\hbox{$\sim$}~}\xspace}

\def\pt         {\ensuremath{p_{\mathrm{T}}}\xspace}

\def\et         {\ensuremath{E_{\mathrm{T}}}\xspace}

\def\bcvegpy    {\mbox{\textsc{Bcvegpy}}\xspace}

\def\pythia     {\mbox{\textsc{Pythia}}\xspace}

\def\tell1  {TELL1\xspace}
\def\ukl1   {UKL1\xspace}

\def\BcBsPi     {{\ensuremath{\B_{c}^{+} \to \B_{s}^{0} \pi^{+} }}\xspace}
\def\BcDKPi     {{\ensuremath{\Bcp \to \Dp \Kp \pim }}\xspace}
\def\BcDsKK     {{\ensuremath{\Bcp \to \Dsp \Kp \Km }}\xspace}
\def\BcDstKPi     {{\ensuremath{\Bcp \to \Dstarp \Kp \pim }}\xspace}

\newcommand{\lhcborcid}[1]{\href{https://orcid.org/#1}{\hspace*{0.1em}\raisebox{-0.45ex}{\includegraphics[width=1em]{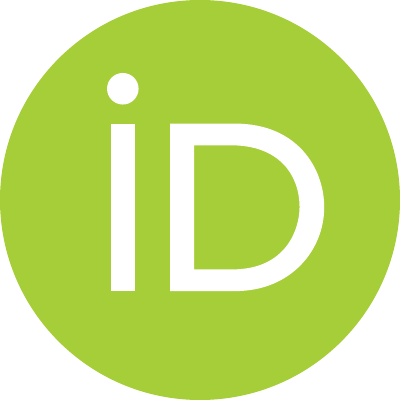}}}}

\usepackage{cite} % Allows for ranges in citations
\usepackage{mciteplus}
\usepackage{longtable} % only for template; not usually to be used in PAPERs

\begin{document}

%%%%%%%%%%%%%%%%%%%%%%%%%
%%%%% Title     %%%%%%%%%
%%%%%%%%%%%%%%%%%%%%%%%%%
\renewcommand{\thefootnote}{\fnsymbol{footnote}}
\setcounter{footnote}{1}

% %%%%%%% CHOOSE TITLE PAGE--------
\onecolumn
% \input{title-LHCb-PAPER}
%%%%%%%%%%% part5 title-LHCb-PAPER.tex
% ===============================================================================
% Purpose: LHCb-PAPER journal paper title page template
% Author: 
% Created on: 2010-09-25
% ===============================================================================

%%%%%%%%%%%%%%%%%%%%%%%%%
%%%%%  TITLE PAGE  %%%%%%
%%%%%%%%%%%%%%%%%%%%%%%%%
\begin{titlepage}
\pagenumbering{roman}

% Header ---------------------------------------------------
\vspace*{-1.5cm}
\centerline{\large EUROPEAN ORGANIZATION FOR NUCLEAR RESEARCH (CERN)}
\vspace*{1.5cm}
\noindent
\begin{tabular*}{\linewidth}{lc@{\extracolsep{\fill}}r@{\extracolsep{0pt}}}
\ifthenelse{\boolean{pdflatex}}% Logo format choice
{\vspace*{-1.5cm}\mbox{\!\!\!\includegraphics[width=.14\textwidth]{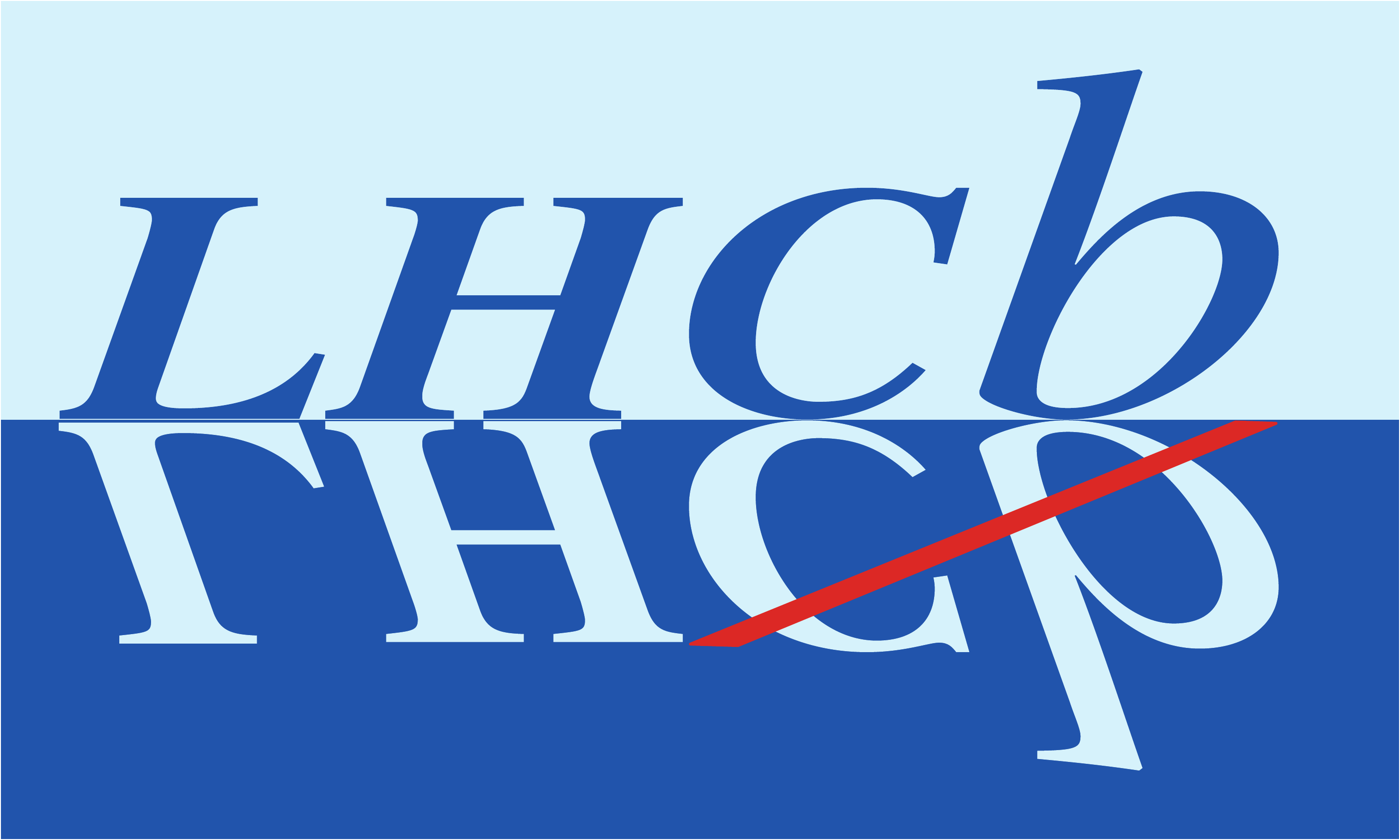}} & &}%
{\vspace*{-1.2cm}\mbox{\!\!\!\includegraphics[width=.12\textwidth]{lhcb-logo.eps}} & &}%
\\
 & & CERN-EP-2025-177 \\  % ID 
 & & LHCb-PAPER-2025-028 \\  % ID 
 & & \today \\ % Date - Can also hardwire e.g.: 23 March 2010
 & & \\
% not in paper \hline
\end{tabular*}

\vspace*{2.0cm}

% Title --------------------------------------------------
{\normalfont\bfseries\boldmath\huge
\begin{center}
% DO NOT EDIT HERE. Instead edit macro in main.tex to keep metadata correct
  \papertitle 
\end{center}
}

\vspace*{2.0cm}

% Authors -------------------------------------------------
\begin{center}
%In the footnote, replace 'paper' by 'Letter' in case of submission to PRL or PLB 
% Edit macro in main.tex to keep metadata correct
\paperauthors\footnote{Authors are listed at the end of this paper.}
\end{center}

\vspace{\fill}

% Abstract -----------------------------------------------
\begin{abstract}
  \noindent
% Searches are presented for $\Bcp$ meson decays into a charm meson and two light pseudoscalar mesons
Searches are presented for $B_{c}^{+} \to D h^+ h^-$ decays, where $D$ is a charmed meson and $h^{\pm}$ is a charged pion or kaon,
using $pp$ collision data collected by the LHCb experiment corresponding to an integrated luminosity of $9~\text{fb}^{-1}$.
The decays $B_c^+\to D^+ K^+\pi^-$, $B_c^+\to D^{*+} K^+\pi^-$ and $B_c^+\to D_s^+ K^+ K^-$ are observed for the first time. Their branching fractions, expressed as ratios relative to that of the $B_c^+\to B_s^0\pi^+$ decay, are determined to be
\begin{align*}
\mathcal{R}(B_c^+\to D^+ K^+\pi^-) =(1.96 \pm 0.23\pm 0.08 \pm 0.10)\times 10^{-3},&\\
\mathcal{R}(B_c^+\to D^{*+} K^+\pi^-) =(3.67 \pm 0.55 \pm 0.24\pm 0.20)\times 10^{-3},&\\
\mathcal{R}(B_c^+\to D_s^+ K^+ K^-) =(1.61 \pm 0.35\pm 0.13\pm 0.07)\times 10^{-3},
\end{align*}
where the first uncertainty is statistical, the second is systematic, and the third is due to the limited precision on the $D$-meson branching fractions.
The decay channels proceed primarily through excited $K^0$ or $D^0$ resonances or $\phi$ mesons, and open a new avenue for studies of charge-parity violation in beauty mesons.
\end{abstract}

\vspace*{1.cm}

\begin{center}
  % Submitted to
  % Phys. Rev. D Lett.
  % Phys.~Rev.~Lett.
  Published in Physical Review Letters 136 (2026) 021804

\end{center}

\vspace{\fill}

{\footnotesize 
% Edit macro in main.tex to keep metadata correct
\centerline{\copyright~\papercopyright. \href{\paperlicenceurl}{\paperlicence}.}}
\vspace*{2mm}

\end{titlepage}

%%%%%%%%%%%%%%%%%%%%%%%%%%%%%%%%
%%%%%  EOD OF TITLE PAGE  %%%%%%
%%%%%%%%%%%%%%%%%%%%%%%%%%%%%%%%

%  empty page follows the title page ----
\newpage
\setcounter{page}{2}
\mbox{~}
%\newpage
%
%% Author List ----------------------------
%%  You need to get a new author list!
%\input{LHCb_authorlist.tex}
%
%The author list for journal publications is provided by the Membership Committee shortly after 'approval to go to paper' has been given.
%%It will be made available on the page
%%\verb!http://www.physik.uzh.ch/~strauman/forMemCo/LHCb-PAPER-XXXX-XXX/! .
%It will be sent to you by email shortly after a paper number has beens assigned.
%The author list should be included already at first circulation, 
%to allow new members of the collaboration to verify whether they have been included correctly.
%Occasionally a misspelled name is corrected or associated institutions become full members.
%In that case, a new author list will be sent to you.
%In case line numbering doesn't work well after including the authorlist, try moving the \verb!\bigskip! after the last author to a separate line.
%
%
%The authorship for Conference Reports should be ``The LHCb
%  collaboration'', with a footnote giving the name(s) of the contact
%  author(s), but without the full list of collaboration names.

%\twocolumn
% %%%%%%%%%%%%% ---------

\renewcommand{\thefootnote}{\arabic{footnote}}
\setcounter{footnote}{0}

%%%%%%%%%%%%%%%%%%%%%%%%%%%%%%%%
%%%%%  Table of Content   %%%%%%
%%%%%%%%%%%%%%%%%%%%%%%%%%%%%%%%
%%%% Uncomment if desired
%\tableofcontents
\cleardoublepage

%%%%%%%%%%%%%%%%%%%%%%%%%
%%%%% Main text %%%%%%%%%
%%%%%%%%%%%%%%%%%%%%%%%%%

\pagestyle{plain} % restore page numbers for the main text
\setcounter{page}{1}
\pagenumbering{arabic}

%% Uncomment during review phase. 
%% Comment before a final submission.
% \linenumbers

%% This is the main body
%% It is useful to have a single file so comemnts are not missed in overleaf.

%%%%%%% part1 body.tex

Decays of $B$ mesons provide a powerful probe for testing the Standard Model (SM) and exploring new physics. 
The $\Bc$ meson,  which is less extensively studied than $\Bz$, $\Bp$ and $\Bs$ mesons, offers a unique opportunity to expand our understanding of $B$-meson properties and explore systems containing two heavy quarks.  
The $\Bc$ meson decays exclusively via weak interactions, through three distinct processes. The decay through a $b$-quark transition contributes about 20\% of the total width, the $c$-quark transition around 70\%, and the remaining 10\% is attributed to the $\bar{b}c\to W^+ \to \bar{q}q'$ weak annihilation amplitude~\cite{Gouz:2002kk}.

Since the discovery of the $\Bc$ meson~\cite{CDF:1998ihx, CDF:1998axz}, the study of its decays has prompted significant theoretical and experimental interest.
The weak decay of the $\Bc$ meson provides an ideal platform for investigating the nonperturbative realm of quantum chromodynamics (QCD) and \CP-violating effects. 
Decays with an open-charm hadron in the final state, $\Bc\to \D X$, are of particular interest because 
they include interfering amplitudes from $\bquarkbar\cquark$ annihilation~\cite{Du:1998te} and $\bquarkbar\to \uquark\uquarkbar \squarkbar (\dquarkbar)$ tree and electroweak loop (penguin) diagrams, which can give rise to observable \CP-violating effects. Charge conjugate processes are implied throughout this Letter.
Branching fractions (BFs) and \CP asymmetries of two-body $\Bc\to \Dss P$ and $\Bc\to \Dss V$ decays, where $P (V)$ represents a pseudoscalar (vector) meson, have been calculated using the perturbative QCD approach~\cite{Rui:2011qc,Zou:2017yxc}. 
However, experimental studies are quite limited due to the relatively low production cross-section of $\Bc$ mesons in $\epem$~\cite{Ai:2024nmn} and $pp$ collisions~\cite{LHCb-PAPER-2019-033} compared to that of \bquark hadrons with lighter quarks.
The first open-charm \Bc decay to be observed was $\Bc \to \Dz \Kp$, reported by the \lhcb collaboration~\cite{LHCb-PAPER-2016-058}. 
This work has been extended in Ref.~\cite{LHCb-PAPER-2024-035} to cover additional studies, including the first observation of the $\Bcp\to\Dp\Kstarz$ decay. The previous BF measurements favor a dominant annihilation contribution in the decay amplitude. These studies offer valuable insights into the contributions of different decay amplitudes in $\Bc$ decays, enabling a better understanding of the SM~\cite{Choi:2009ym,Zhu:2018epc}. 
Further experimental results are needed to understand how the different quark-level processes contribute in different decay channels.

\begin{figure}[t]
    \centering
    \includegraphics[width = 0.45 \columnwidth]{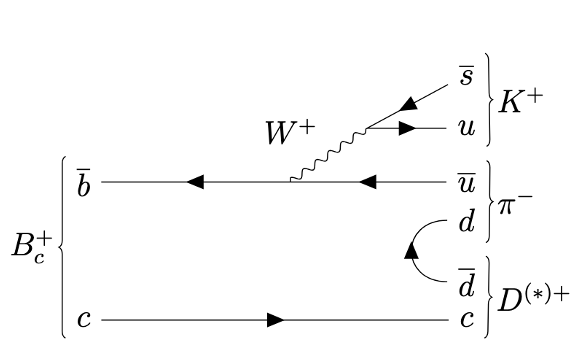}
    \includegraphics[width = 0.45 \columnwidth]{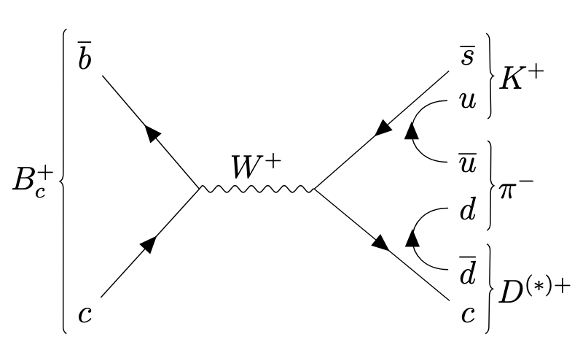}
    \includegraphics[width = 0.45 \columnwidth]{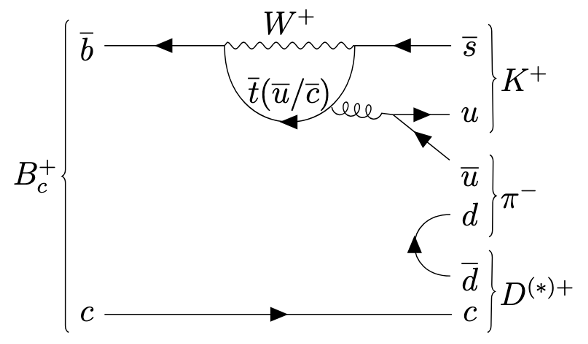}
    \caption{Possible Feynman diagrams for the $B_c^+ \to D^{(*)+} K^+ \pi^-$ decay.}
    \label{fig:feynman}
\end{figure}

This letter describes searches for the decays $\BcDKPi$, $\BcDstKPi$ and $\BcDsKK$, which proceed via processes such as those depicted in Fig.~\ref{fig:feynman}. 
These three-body $\Bc$ decays are expected to exhibit a rich dynamic structure with multiple intermediate resonances. 
Amplitude analyses of multibody decays, with their rich interference patterns and sensitivity to phases, provide unique insights into resonance properties, polarization fractions, and \CP\ violation.
The measurements described here are based on $pp$ collision data collected by the \lhcb experiment at center-of-mass energies of 7, 8, and 13\tev, corresponding to a total integrated luminosity of about $9\invfb$. 

The \lhcb detector~\cite{LHCb-DP-2008-001,LHCb-DP-2014-002,Aaij:2014zzy,LHCbOuterTrackerGroup:2013epe,Alves:2012ey} is a single-arm forward spectrometer covering the pseudorapidity range $2 < \eta < 5$, designed for the study of particles containing $b$ or $c$ quarks. 
The detector elements that are particularly relevant to this analysis are: a silicon-strip vertex detector surrounding the $pp$ interaction region that allows $c$ and $b$ hadrons to be identified from their characteristically long flight distance; a tracking system that provides a measurement of the momentum, $p$, of charged particles; and two ring-imaging Cherenkov (RICH) detectors that are able to discriminate between different species of charged hadrons.
% described in detail in Refs.
Photons, electrons and hadrons are identified by a calorimeter system consisting of
scintillating-pad and preshower detectors, an electromagnetic
% calorimeter
and a hadronic calorimeter~\cite{LHCb-DP-2020-001}. Muons are identified by a
system composed of alternating layers of iron and multiwire
proportional chambers~\cite{Alves:2012ey}.
The online event selection is performed by a trigger~\cite{LHCb-DP-2012-004}, 
which consists of a hardware stage~\cite{LHCb-DP-2019-002}, based on information from the calorimeter and muon
systems, followed by a software stage, which applies a full event
reconstruction.
Simulation samples are generated using \bcvegpy implemented in \pythia~8~\cite{Sjostrand:2007gs,Chang:2003cq,Chang:2005hq,Wang:2012ah,Wu:2013pya} with a dedicated LHCb configuration~\cite{LHCb:2011dpk}. These samples play an important role in optimizing the event selection and estimating the selection efficiencies of signal decays.

The $D$-meson candidates are reconstructed from their decays $\Dp \to \Km \pip \pip$, \mbox{$\Dstarp\to\Dz(\to\Km\pip)\pip$} and $\Dsp\to\Kp\Km\pip$, which have relatively large BFs. These $D$-meson candidates are subsequently combined with charged pions and kaons to form $\Bc$ candidates. The known $\BcBsPi$ decay, with $\Bs \to \Dsm(\to\Kp\Km\pip) \pip$, is used as a normalization channel, enabling measurements of the BF ratios
\begin{equation}
\label{eq:eff-intro}
    \begin{aligned}
        \mathcal{R}_{\Bcp \to D h^{+} h^{-}} 
        & \equiv \frac{\mathcal{B}(\Bcp \to D h^{+} h^{-})}{\mathcal{B}(\BcBsPi)} \\
        &= \frac{N_{\Bcp \to D h^{+} h^{-}}}{N_{\BcBsPi}} \cdot \frac{\varepsilon_{\BcBsPi}}{\varepsilon_{\Bcp \to D h^{+} h^{-}}} \cdot \frac{\mathcal{B}(\Bs)}{\mathcal{B}(D_{(s)}^{(*)+})},
    \end{aligned}
\end{equation}
where $N_{\Bc \to X}$ represents the yield of $\Bc \to X$ decays, $\varepsilon_{\Bc \to X}$ represents their selection efficiency, $\mathcal{B}(D_{(s)}^{(*)+})$ represents the appropriate $D$-meson branching fraction for the signal channels, and $\mathcal{B}(\Bs)$ represents the BF product $\mathcal{B}(\Bs \to \Dsm \pip)\cdot \mathcal{B}(\Dsm \to \Kp \Km \pim)$.

At  the hardware-trigger level, events events are required to have a hadron with  transverse energy $\et$ in the calorimeters exceeding 3.5~\gev~\cite{LHCb-DP-2020-001}.
These trigger objects may originate directly from the $\Bc$ decays of interest or from other particles in the event. The subsequent software trigger performs full event reconstruction and imposes stricter requirements: at least one charged particle must possess higher transverse momentum ($\pt$) than the hardware threshold and be significantly displaced from all reconstructed primary interaction vertices (PVs). Furthermore, the reconstructed events must have signal candidates forming a two-, three-, or four-track secondary vertex with a significant displacement from any PV~\cite{Gligorov:2012qt,LHCb-PROC-2015-018}.
For the offline event selection, momentum requirements are applied to reconstructed tracks to ensure that they fall within the kinematic ranges where the RICH detectors operate at full efficiency~\cite{LHCbRICHGroup:2012mgd}. Additional requirements, including on particle identification (PID) quantities, track-fit quality, and the minimum impact parameter of a track relative to any PV, are applied to all final-state particles to ensure consistency with the signal decays.
Masses of the reconstructed $D$ candidates are required to be within $\pm20\mevcc$ (for $\Dp$ and $\Dsp$ candidates) and $\pm2\mevcc$ (for $\Dstarp$ candidates) of the known values~\cite{PDG2024}.  A kinematic fit is performed to each candidate~\cite{Hulsbergen:2005pu}, with vertex constraints applied to both the $B$ and $D$ vertices, and with the $D$ candidates' ($\Dp$, $\Dz$ and $\Ds$) masses constrained to their known values~\cite{PDG2024}. This has the effect of significantly improving the \Dstarp mass resolution.

\begin{figure}[tb]
    \centering
    \includegraphics[width = 0.45 \columnwidth]{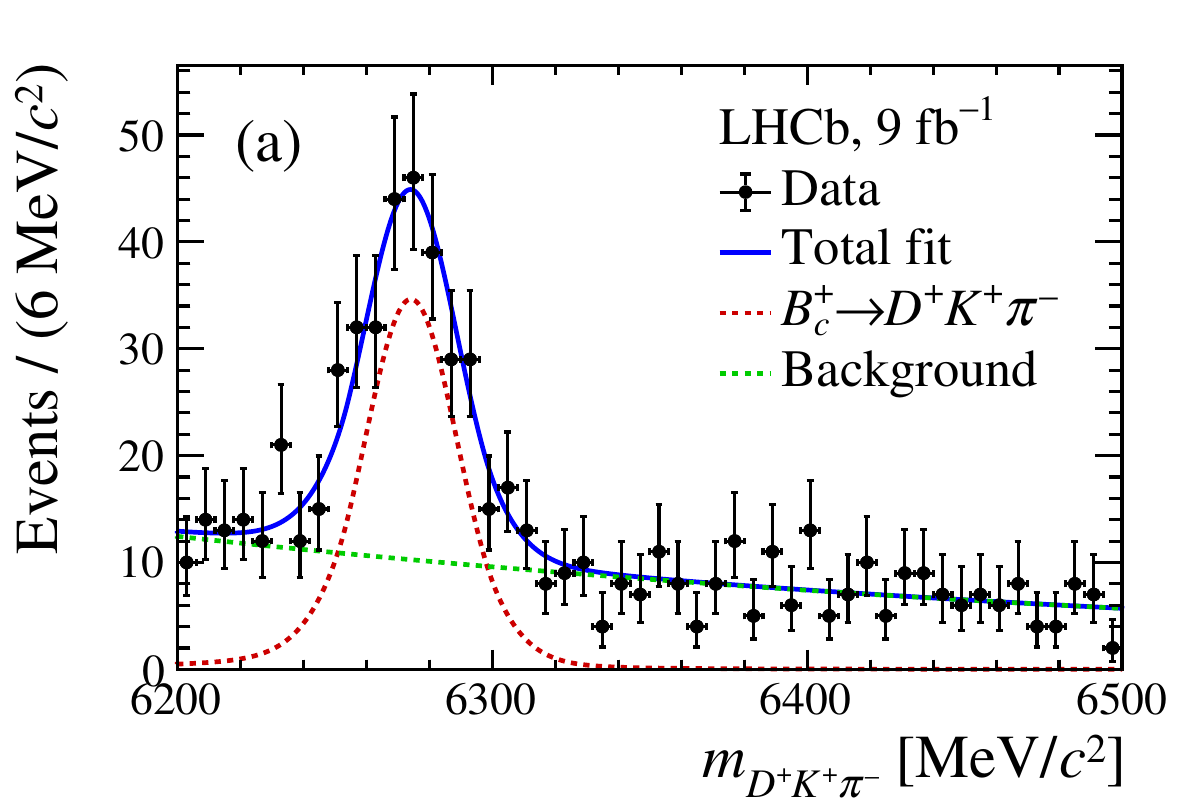}
    \includegraphics[width = 0.45 \columnwidth]{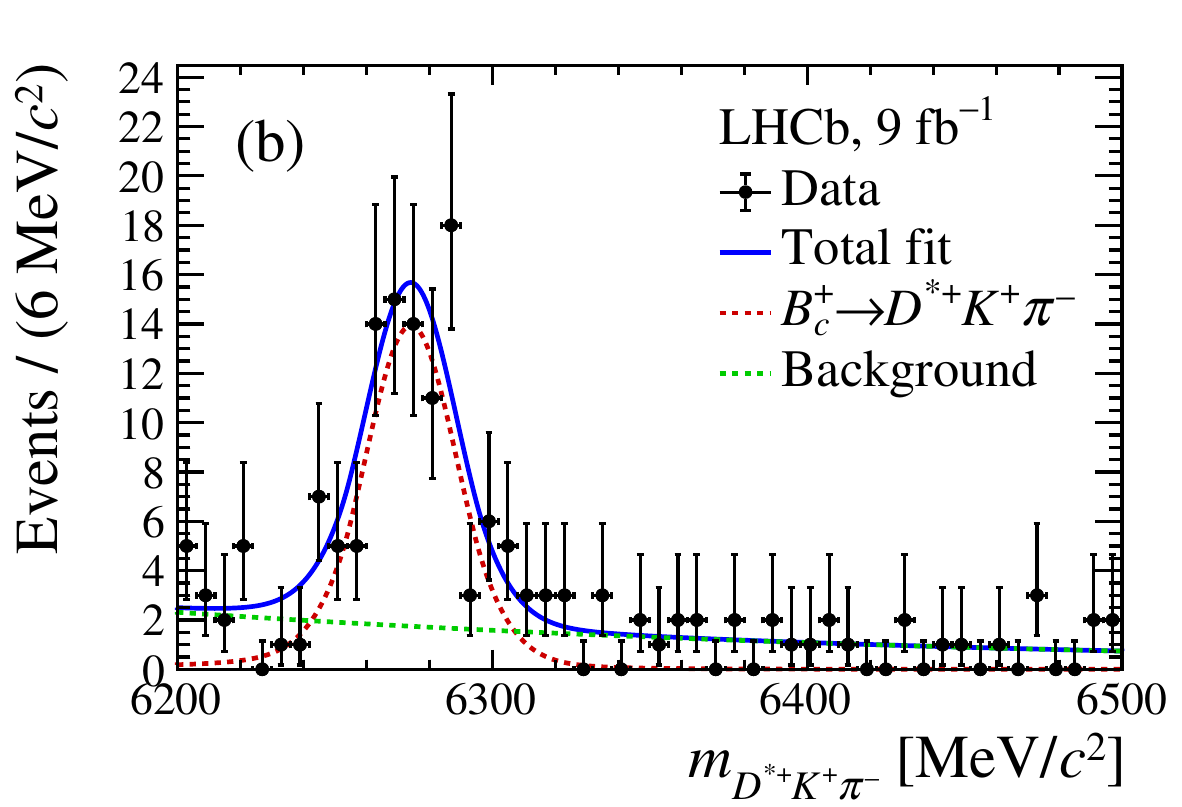}
    \includegraphics[width = 0.45 \columnwidth]{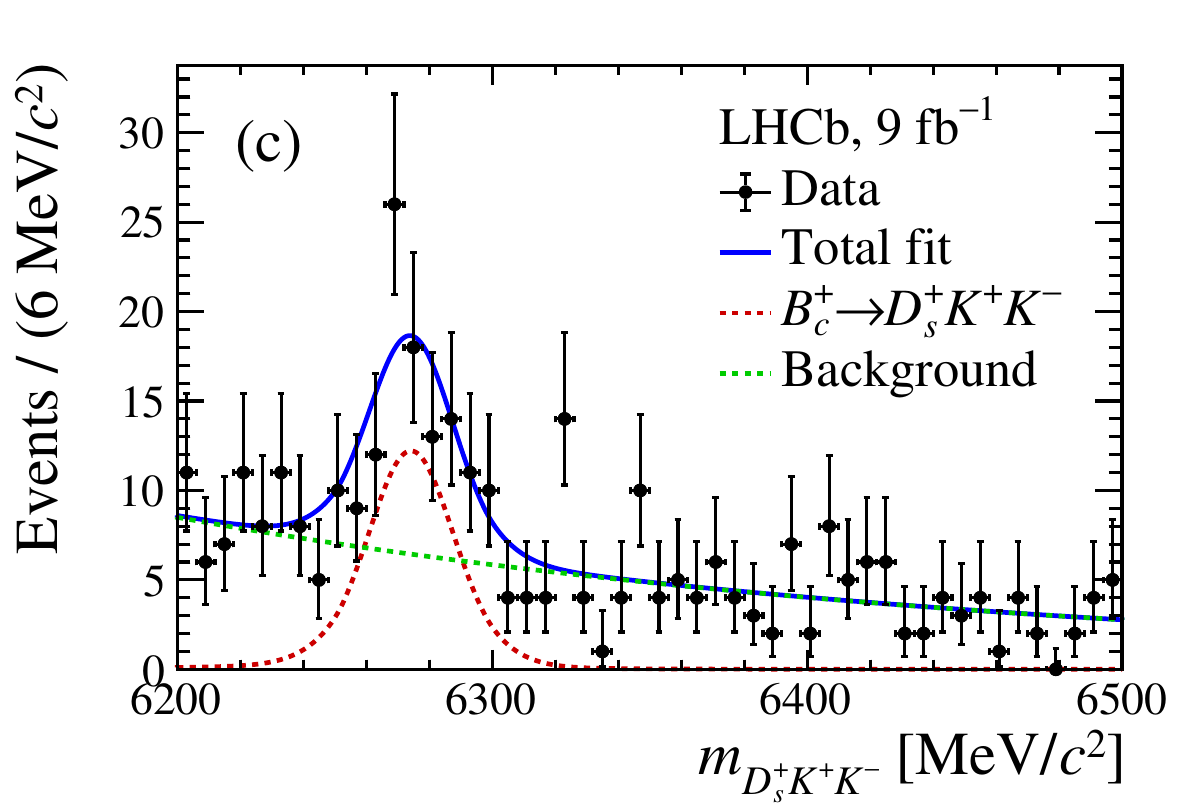}
    \includegraphics[width = 0.45 \columnwidth]{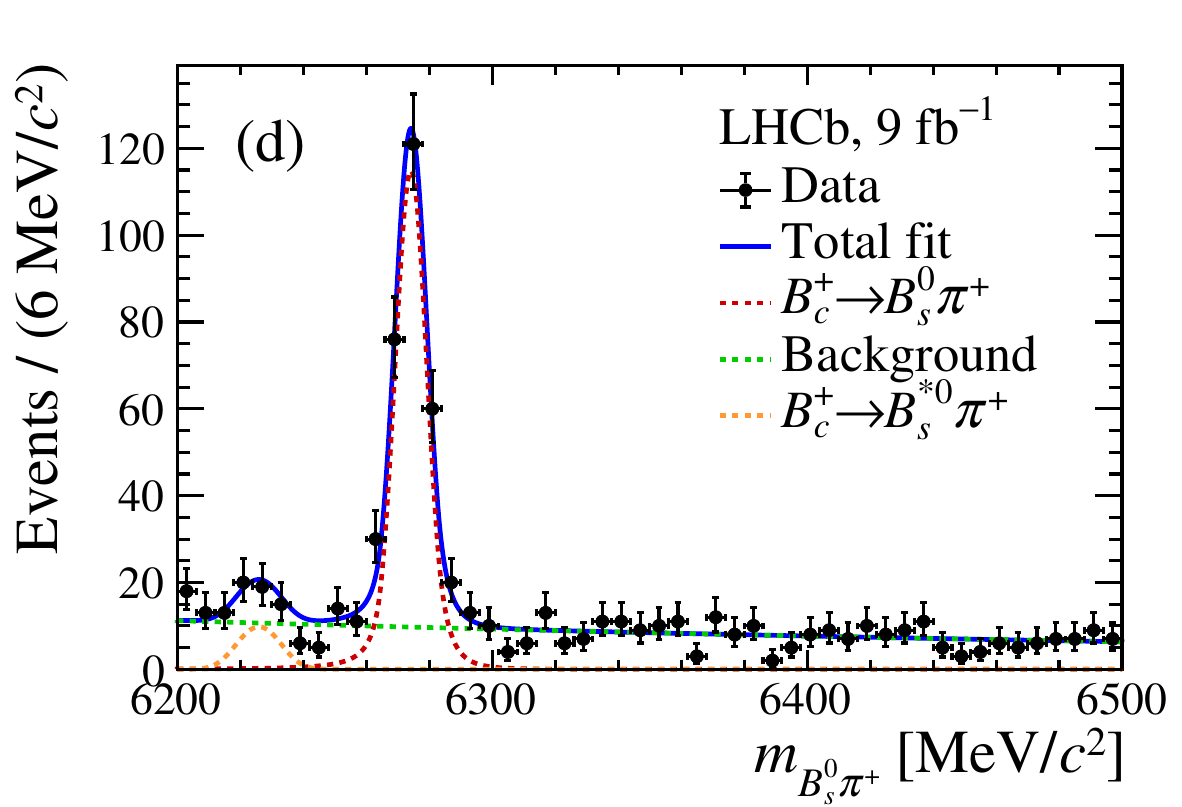}
    \caption{Mass distributions for (a) $B_c^+\to D^+ K^+\pi^-$, (b) $B_c^+\to D^{*+} K^+\pi^-$, (c) $B_c^+\to D_s^+ K^+ K^-$ and (d) $B_c^+\to B_s^0\pi^+$ candidates, after all selection criteria are imposed. The fit results are also shown.}
    \label{fig:signal_fit}
\end{figure}

To further suppress background, a multivariate analysis based on a boosted decision tree (BDT) classifier~\cite{Roe:2004na} is implemented for each channel. The classifier is trained using a set of observables selected for their discriminating power between signal and background candidates. The input variables for the BDT training include calibrated PID quantities and topological observables, such as the impact-parameter significance, flight distance with respect to the PV, and the vertex-fit $\chi^2$, for different particles.
For the training, simulated signal candidates are used as a proxy for the true signal while the background proxy sample uses data from the high-$\Bc$-mass sideband region $[6450,6750]\mevcc$.
To mitigate the effects of statistical fluctuations, the k-fold method~\cite{kFold} is applied to the samples during training. 
The requirement on the BDT output is chosen to maximize the figure of merit, $\varepsilon_{s}/\sqrt{S+B}$, where $\varepsilon_{s}$ is the signal efficiency estimated from simulation, $S$ is the signal yield estimated based on the assumption $\mathcal{B}(\BcDKPi)=2\cdot\mathcal{B}(\Bc\to\Dp\Kstar(892)^{0}(\to \Kp\pim))$, and $B$ is the background yield in the signal region (defined as being within $\pm 40 \mevcc$ of the known $\Bc$ mass) estimated from the sideband sample. The factor 2 in the assumption takes the other excited $\Kstar$ states into consideration. The assumption  $\mathcal{B}(\BcDstKPi)=\mathcal{B}(\BcDsKK)=\mathcal{B}(\BcDKPi)$ is made when optimizing the BDT requirement for the other two signal decays. 
The resulting mass spectra for the $\Dp\Kp\pim$, $\Dstarp\Kp\pim$, $\Dsp\Kp\Km$ and $\Bs\pip$ final states, after applying all selection criteria, are shown in Fig.~\ref{fig:signal_fit}.

A simultaneous unbinned maximum-likelihood fit is performed to all four mass spectra to determine the yield of each decay. To minimize the influence of partially reconstructed background in the signal channels, the mass fit range is chosen to be [6200, 6500]$\mevcc$. 
For all decays, the signal shape is described using the sum of a Gaussian function and a Crystal Ball function~\cite{Skwarnicki:1986xj}, modified to include power-law tails on both sides of the Gaussian core.
The power-law tail parameters are fixed to values obtained from fits to the simulation. The four decay channels share a common \Bc peak position and width ratio for the two peaking functions, which are fixed to values obtained from simulation, allowing only one width to float. The main background contribution is combinatorial in nature, and is described by an exponential function. For the normalization channel $\BcBsPi$, the sample is contaminated by the decay $\Bc\to B^{*0}_{s}(\to \Bs \gamma)\pip$ where the photon is not reconstructed. This component is described by a shape obtained from the fast simulation package RapidSim~\cite{Cowan:2016tnm}. The signal and background yields are left unconstrained during the fit. The fit projections are shown in Fig.~\ref{fig:signal_fit}.  

The extracted signal yields are $230\pm23$ for the $\BcDKPi$ decay, $87\pm12$ for the $\BcDstKPi$ decay, $68\pm13$ for the $\BcDsKK$ decay and $267\pm19$ for the $\BcBsPi$ decay, where the uncertainties are statistical only. The signal significances are evaluated using Wilks' theorem~\cite{Wilks:1938dza}, which is derived from the likelihood difference between the background-only and signal-plus-background hypotheses. 
The signal significance for all decay modes exceeds five standard deviations, after accounting for systematic uncertainties due to the choice of fit model.

\begin{figure}[h]
    \centering
    \includegraphics[width = 0.45 \columnwidth]{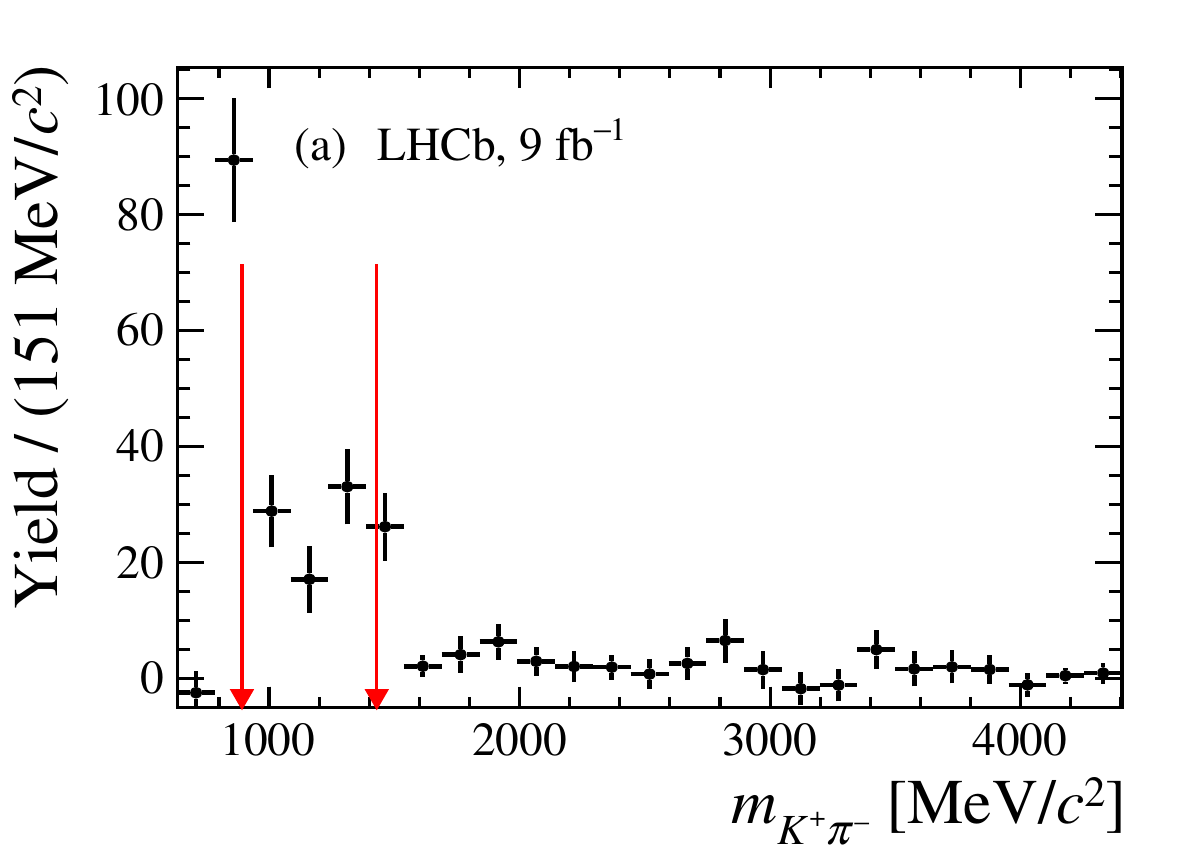}
    \includegraphics[width = 0.45 \columnwidth]{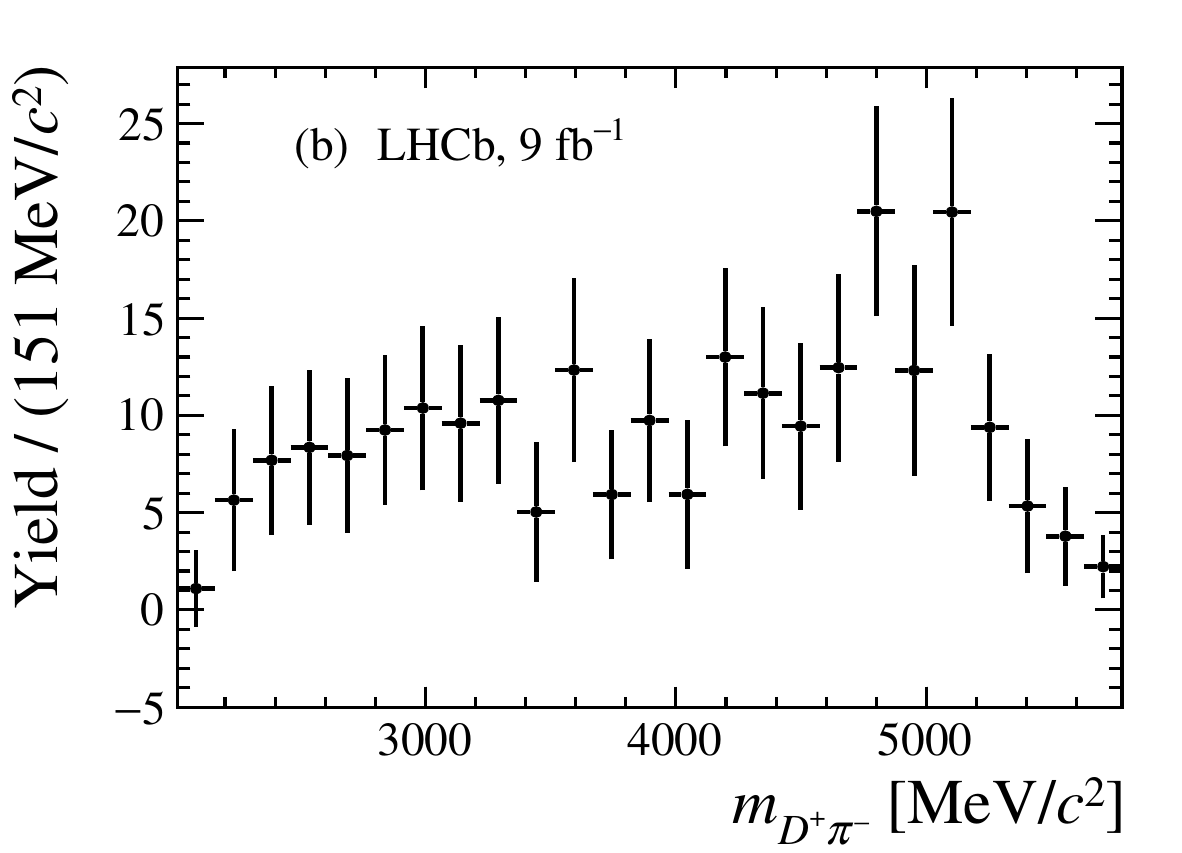}\\
    \includegraphics[width = 0.45 \columnwidth]{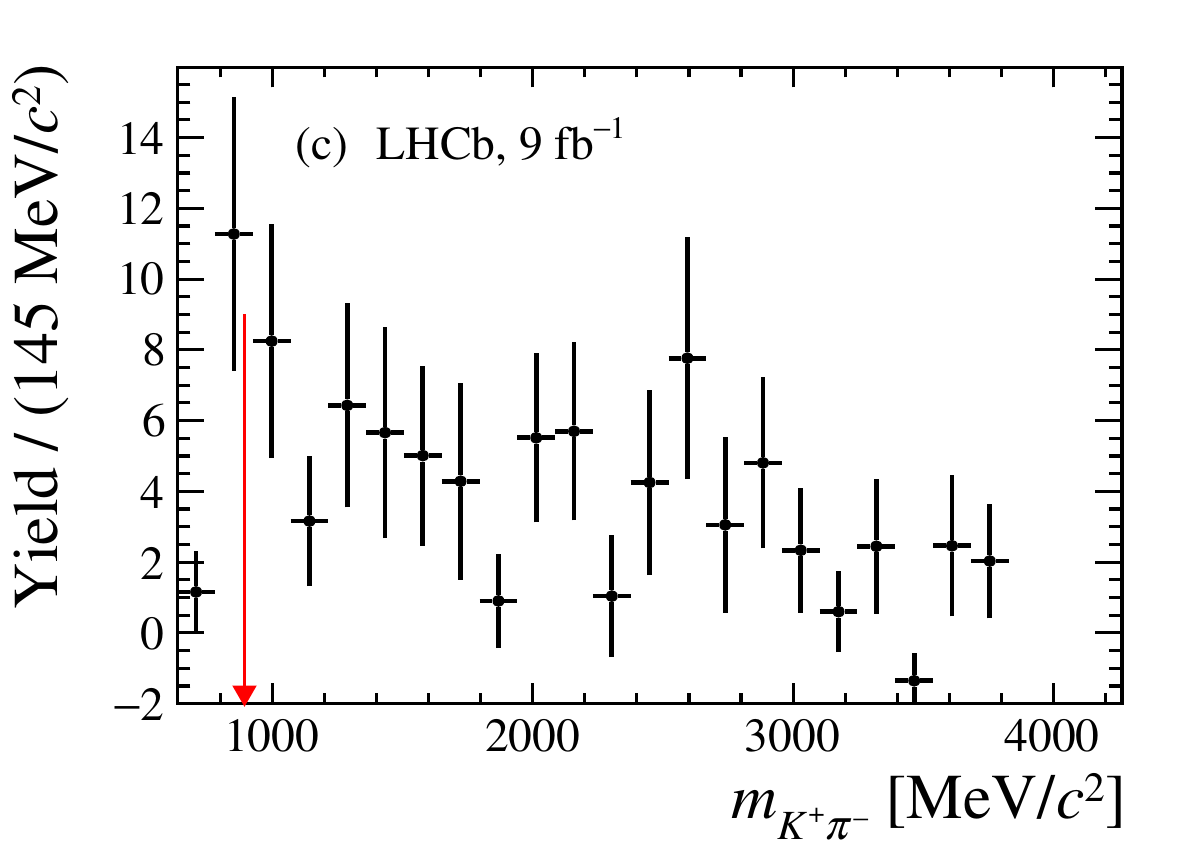}
    \includegraphics[width = 0.45 \columnwidth]{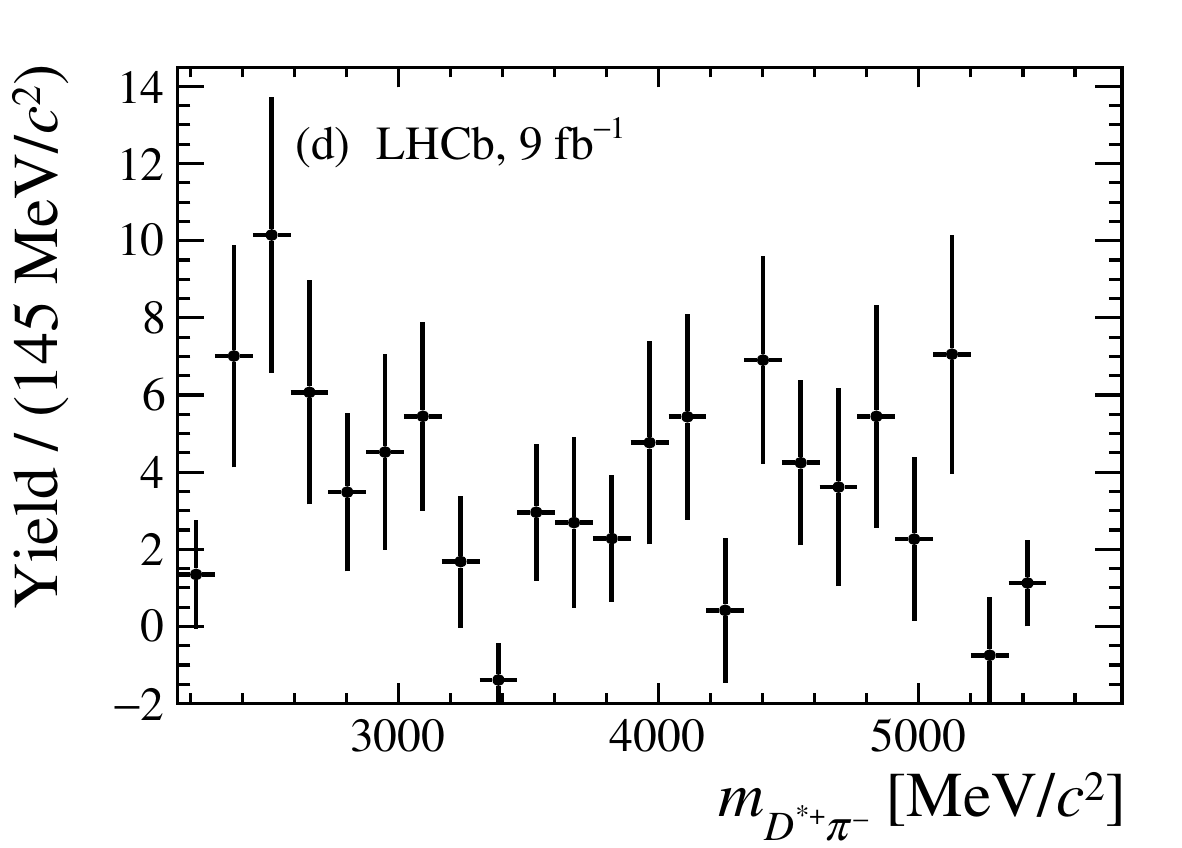}\\
    \includegraphics[width = 0.45 \columnwidth]{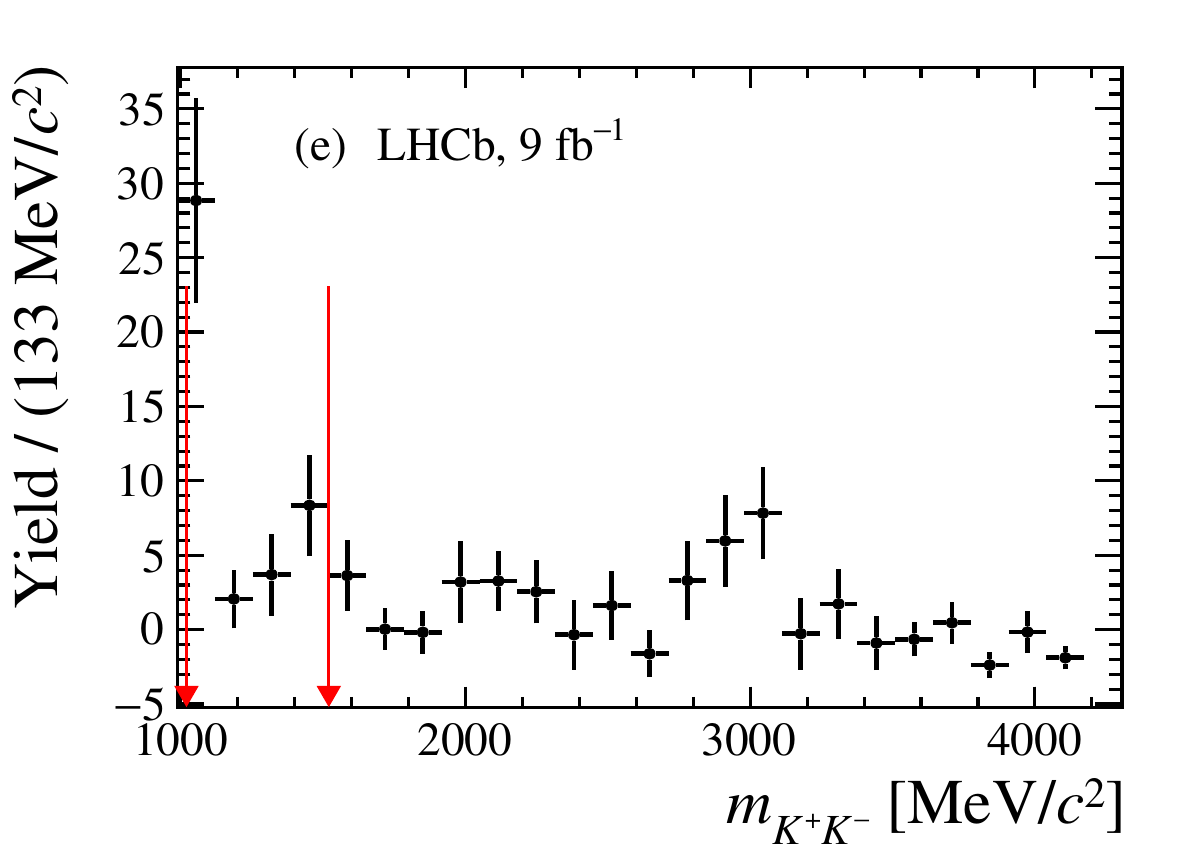}
    \includegraphics[width = 0.45 \columnwidth]{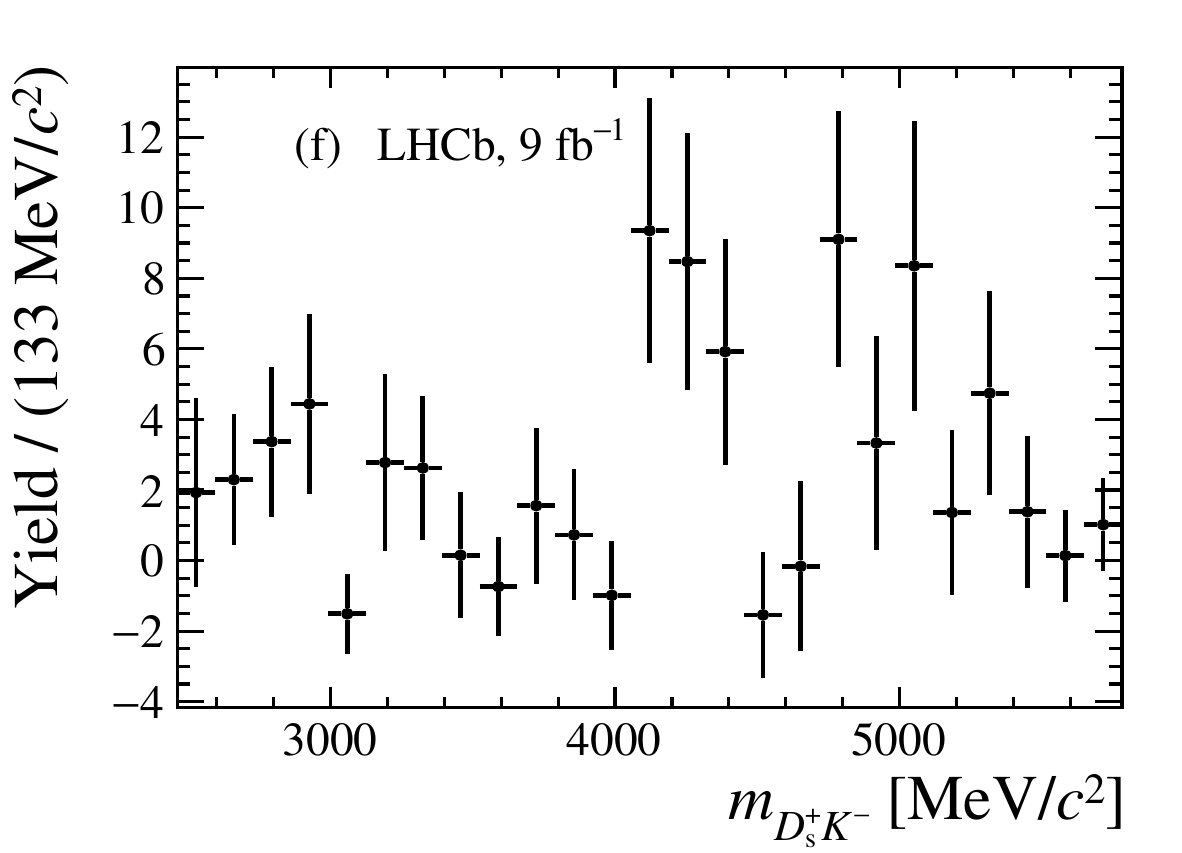}

    \caption{Two-body mass distributions for (a, b) $B_c^+\to D^+ K^+\pi^-$, (c, d) $B_c^+\to D^{*+} K^+\pi^-$, and (e, f) $B_c^+\to D_s^+ K^+ K^-$ decays. The red arrows denote the positions of possible resonances in the $K^+ K^-$ or $K^+ \pi^-$ systems.}
    \label{fig:2body}
\end{figure}

The background-subtracted~\cite{Pivk:2004ty,Xie:2009rka} two-body mass distributions of signal decays are shown in Fig.~\ref{fig:2body}. 
A significant peak compatible with the $\Kstar(892)^0$ state and a smaller peak compatible with the $\Kstar(1430)^{0}$ state are seen in the $m_{\Kp\pim}$ mass spectrum for the $\BcDKPi$ decay. A peak compatible with the $\Kstar(892)^{0}$ state is also visible in the $m_{\Kp\pim}$ mass spectrum for the $\BcDstKPi$ decay. A significant contribution in the low-mass $m_{\Kp\Km}$ region is observed in the $\BcDsKK$ decay, with part of it originating from the $\phi(1020)$ resonance. Additionally, a hint of the $f_{2}^{'}(1525)$ state may also be present. 

Measurements of the branching fractions require corrections for the efficiency to detect, reconstruct, and select the decays under study.  For the three-body decays, the efficiency distribution across the phase space is nonuniform and is therefore calculated as a function of the two-dimensional mass combinations  ($m_{D^{(*)+}_{(s)}K^+},m_{D^{(*)+}_{(s)}h^-}$). The total efficiency is calculated as the product of several components, including the geometrical acceptance and reconstruction, trigger and selection efficiencies,  all of which are derived from simulation samples. To ensure good agreement between data and simulation, various corrections are applied to the simulated samples. The PID variables in the simulation are calibrated using data from $\Dstarp \to \Dz(\to \Km \pip) \pip$ decays~\cite{Aaij:2018vrk}.
The $\pt$ distribution of $\Bcp$ mesons in the simulation sample are weighted to match those in the data sample, using the $\Bc\to J/\psi \pip$ decay, which has a large signal yield. Due to inaccuracies in the simulation of the hardware trigger, the trigger efficiency is corrected using the data-driven method of Ref.~\cite{LHCb-DP-2012-004} to align the efficiency in the simulation with that in
the data.

\begin{table}[tb]
\centering
\caption{Absolute systematic uncertainties (in units of $10^{-3}$) on the BF ratio measurements.}
\renewcommand{\arraystretch}{1.1}
\resizebox{1\columnwidth}{!}{
\begin{tabular}{cccc}
\hline\hline
Sources   & $\mathcal{R}_{\BcDKPi}$ & $\mathcal{R}_{\BcDstKPi}$ & $\mathcal{R}_{\BcDsKK}$ 
\\ \hline
Fit model & 0.02 & 0.05 & 0.02 \\ 
Simulation sample size & 0.04 & 0.09 & 0.06 \\
$\pt$ weight & 0.01 & 0.09 & 0.03 \\
2D efficiency binning & 0.06 & 0.18 & 0.08 \\
Trigger correction & 0.01 & 0.02 & 0.01 \\
PID calibration & 0.02 & 0.07 & 0.05 \\ \hline
Total & 0.08 & 0.24 & 0.13 \\
\hline \hline
\end{tabular}
}
\label{tab:sys}
\end{table}

Many systematic effects cancel out in the BF ratio due to similar decay topology of signal and normalization channels. The remaining sources of systematic uncertainty in the BF ratio measurement are summarized in Table~\ref{tab:sys}. The total systematic uncertainty is obtained by summing all contributions in quadrature. 
To account for the effect of the particular signal and background models used in the mass fit, alternative descriptions are used. These are a Cruijff function~\cite{BaBar:2010vgu} for the signal peaks and a first-order Chebyshev polynominal function for the background. The resulting shift in the measurement with respect to the baseline result is assigned as a systematic uncertainty. 
The uncertainty due to the finite simulation sample size is included as a subdominant systematic uncertainty in the BF ratio measurement.
The $\pt$ weighting procedure, calibrated using $\Bc \to J/\psi \pip$ decays, is applied to the simulation samples. Due to the limited sample size of the calibration sample, this introduces a systematic uncertainty. To estimate its impact on the BF measurement, the uncertainty of the $\pt$ weights are propagated to the BF measurement using pseudoexperiments.

The efficiency is determined as a function of ($m_{D^{(*)+}_{(s)}K^+},m_{D^{(*)+}_{(s)}h^-}$). Variations in the binning scheme used for this evaluation lead to different results. To quantify the associated uncertainty, different binning schemes ($10\times 10$, $15\times 15$, $20\times 20$ and $25\times 25$) are applied when calculating the BF ratios, and the maximum variation is assigned as a systematic uncertainty.
The systematic uncertainty from the hardware trigger correction, is determined using the method of Ref.~\cite{LHCb-DP-2012-004}.
The PID variables from the simulated samples are corrected to match the large high-purity calibration samples~\cite{LHCb-PUB-2016-021}. The associated uncertainty is estimated by using alternative PID templates. 

Taking all systematic uncertainties into account, the BF ratio between each signal decay and the normalization decay are determined by Eq.~\ref{eq:eff-intro} to be 
\begin{equation*}
    \begin{split}
        \mathcal{R}_{\BcDKPi} &= (1.96 \pm 0.23 \pm 0.08 \pm 0.10) \times 10^{-3}, \\
        \mathcal{R}_{\BcDstKPi} &= (3.67 \pm 0.55 \pm 0.24 \pm 0.20) \times 10^{-3}, \\
        \mathcal{R}_{\BcDsKK} &= (1.61 \pm 0.35 \pm 0.13 \pm 0.07) \times 10^{-3}.
    \end{split}
\end{equation*}
where the first uncertainty is statistical, the second is systematic, and the third is due to the limited precision on the $D$-meson branching fractions~\cite{PDG2024}.

To summarize, the decays $\BcDKPi$, $\BcDstKPi$ and $\BcDsKK$ are observed for the first time, using a data sample corresponding to an integrated luminosity of 9\invfb collected by the LHCb experiment. Their branching fractions relative to the $\BcBsPi$ decay are measured.
Following the recent upgrade of the LHCb detector~\cite{LHCb-DP-2022-002} , the experiment is now entering a new phase of data collection characterised by higher luminosity and enhanced trigger capabilities. This development opens the door to unprecedented investigations into hadronic \Bc decays.
This work lays a crucial groundwork for future searches for local \CP violation in \Bc decays, offering valuable insights into the underlying quark-level dynamics and contributing to our broader understanding of \CP phenomena within the Standard Model.

% Do not include this in any draft (just for information in the template)

%%%%%%% part2 acknowledgements.tex
\emph{Data availability:} Data associated to the plots in this publication are made available on the CERN Document Server in Ref.~\cite{LHCb-PAPER-2025-028-cds}.
\section*{Acknowledgements}
%
% These Acknowledgements valid from 3-May-2019
%
\noindent We express our gratitude to our colleagues in the CERN
accelerator departments for the excellent performance of the LHC. We
thank the technical and administrative staff at the LHCb
institutes.
We acknowledge support from CERN and from the national agencies:
ARC (Australia);
CAPES, CNPq, FAPERJ and FINEP (Brazil); 
MOST and NSFC (China); 
CNRS/IN2P3 (France); 
BMFTR, DFG and MPG (Germany); 
INFN (Italy); 
NWO (Netherlands); 
MNiSW and NCN (Poland); 
MCID/IFA (Romania); 
%MSHE (Russia); 
MICIU and AEI (Spain);
SNSF and SER (Switzerland); 
NASU (Ukraine); 
STFC (United Kingdom); 
DOE NP and NSF (USA).
%%%%%%%%%%%%%%%%%%%%%%%%%%%%%%%%%%%%%%%%%%%%%
We acknowledge the computing resources that are provided by ARDC (Australia), 
CBPF (Brazil),
CERN, 
IHEP and LZU (China),
IN2P3 (France), 
KIT and DESY (Germany), 
INFN (Italy), 
SURF (Netherlands),
Polish WLCG (Poland),
IFIN-HH (Romania), 
%RRCKI and Yandex LLC (Russia), 
PIC (Spain), CSCS (Switzerland), 
and GridPP (United Kingdom).
%%%%%%%%%%%%%%%%%%%%%%%%%%%%%%%%%%%%%%%%%%
We are indebted to the communities behind the multiple open-source
software packages on which we depend.
%%%%%%%%%%%%%%%%%%%%%%%%%%%%%%%%%%%%%%%%%%
Individual groups or members have received support from
%ARC and ARDC (Australia); % moved to national 16/01
Key Research Program of Frontier Sciences of CAS, CAS PIFI, CAS CCEPP, 
Fundamental Research Funds for the Central Universities,  and Sci.\ \& Tech.\ Program of Guangzhou (China);
Minciencias (Colombia);
EPLANET, Marie Sk\l{}odowska-Curie Actions, ERC and NextGenerationEU (European Union);
A*MIDEX, ANR, IPhU and Labex P2IO, and R\'{e}gion Auvergne-Rh\^{o}ne-Alpes (France);
%RFBR, RSF and Yandex LLC (Russia);
Alexander-von-Humboldt Foundation (Germany);
ICSC (Italy); 
%GVA, XuntaGal, GENCAT, Inditex, InTalent and Prog.~Atracci\'on Talento, CM (Spain);
Severo Ochoa and Mar\'ia de Maeztu Units of Excellence, GVA, XuntaGal, GENCAT, InTalent-Inditex and Prog.~Atracci\'on Talento CM (Spain);
SRC (Sweden);
the Leverhulme Trust, the Royal Society and UKRI (United Kingdom).

% Comment this in for paper drafts; do not include this in analysis note, conference and figure reports
% \input{supplementary}

\clearpage
\addcontentsline{toc}{section}{References}
%\setboolean{inbibliography}{true}
\bibliographystyle{LHCb}
% \bibliography{main,standard,LHCb-PAPER,LHCb-CONF,LHCb-DP,LHCb-TDR}
%%%%%%% part7 bib
\ifx\mcitethebibliography\mciteundefinedmacro
\PackageError{LHCb.bst}{mciteplus.sty has not been loaded}
{This bibstyle requires the use of the mciteplus package.}\fi
\providecommand{\href}[2]{#2}

\newpage
% \input{Authorship_LHCb-PAPER-2025-028}
%%%%%%%%%%% part6 Authorship_LHCb-PAPER-2025-028.tex

% LHCb collaboration author list
% Data extracted on September 15th, 2025 at 3:25am for paper reference LHCb-PAPER-2025-028
\centerline
{\large\bf LHCb collaboration}
\begin
{flushleft}
\small
R.~Aaij$^{38}$\lhcborcid{0000-0003-0533-1952},
A.S.W.~Abdelmotteleb$^{57}$\lhcborcid{0000-0001-7905-0542},
C.~Abellan~Beteta$^{51}$\lhcborcid{0009-0009-0869-6798},
F.~Abudin{\'e}n$^{57}$\lhcborcid{0000-0002-6737-3528},
T.~Ackernley$^{61}$\lhcborcid{0000-0002-5951-3498},
A. A. ~Adefisoye$^{69}$\lhcborcid{0000-0003-2448-1550},
B.~Adeva$^{47}$\lhcborcid{0000-0001-9756-3712},
M.~Adinolfi$^{55}$\lhcborcid{0000-0002-1326-1264},
P.~Adlarson$^{85}$\lhcborcid{0000-0001-6280-3851},
C.~Agapopoulou$^{14}$\lhcborcid{0000-0002-2368-0147}
% ...

{\footnotesize \it

% ...
$^{14}$Universit{\'e} Paris-Saclay, CNRS/IN2P3, IJCLab, Orsay, France\\
% ...
$^{38}$Nikhef National Institute for Subatomic Physics, Amsterdam, Netherlands\\
% ...
$^{47}$Instituto Galego de F{\'\i}sica de Altas Enerx{\'\i}as (IGFAE), Universidade de Santiago de Compostela, Santiago de Compostela, Spain\\
% ...
$^{51}$Physik-Institut, Universit{\"a}t Z{\"u}rich, Z{\"u}rich, Switzerland\\
% ...
$^{55}$H.H. Wills Physics Laboratory, University of Bristol, Bristol, United Kingdom\\
% ...
$^{57}$Department of Physics, University of Warwick, Coventry, United Kingdom\\
% ...
$^{60}$School of Physics and Astronomy, University of Glasgow, Glasgow, United Kingdom\\
$^{61}$Oliver Lodge Laboratory, University of Liverpool, Liverpool, United Kingdom\\
% ...
$^{69}$Syracuse University, Syracuse, NY, United States\\
% ...
$^{85}$Department of Physics and Astronomy, Uppsala University, Uppsala, Sweden, associated to $^{60}$\\
% ...
}
\end{flushleft}

\end{document}